\documentstyle[epsf,psfig]{l-aa}
\def\CMB{colour-magnitude band }

\thesaurus{11.03.4 Coma; 11.03.5; 11.12.2; 11.16.1}

\begin{document}
\title{Environmental effects on the Coma cluster luminosity function
\thanks{Based on observations collected at 
the Canada-France-Hawaii telescope, operated by the 
National Research Council of Canada, the Centre National de la 
Recherche
Scientifique of France, and the University of Hawaii}}

\author {
C.~Lobo \inst{1,2}
\and
 A.~Biviano \inst{3, 4}
\and
  F.~Durret \inst{1,5}
\and
  D.~Gerbal \inst{1,5}  
\and
 O.~Le F\`evre \inst{5}
\and
 A.~Mazure \inst{6} 
\and
 E.~Slezak \inst{7}
}
\offprints{C.~Lobo, lobo@iap.fr }
\institute{
	Institut d'Astrophysique de Paris, CNRS, Universit\'e Pierre et 
	Marie Curie, 98bis Bd Arago, F-75014 Paris, France 
\and
	Centro de Astrof\'\i sica da Universidade do Porto, Rua do Campo 
	Alegre 823, P-4150 Porto, Portugal
\and
	Istituto T.E.S.R.E., CNR, via Gobetti 101, I-40129 Bologna, Italy
\and
	ESA Villafranca Satellite Tracking Station, ISO Science Team, 
	Apto. 50727, CAM IDT, E-28080 Madrid, Spain
\and 
	DAEC, Observatoire de Paris, Universit\'e Paris VII, CNRS (UA 173), 
	F-92195 Meudon Cedex, France 
\and
    	IGRAP, LAS, Traverse du Siphon, Les Trois Lucs, B.P. 8, F-13376 
	Marseille Cedex, France 
\and
    	Observatoire de la C\^ote d'Azur, B.P. 229, F-06304 Nice Cedex 4, 
	France 
}
\date{Received, 1996 ; accepted,}

\maketitle

\begin{abstract}
Using our catalogue of V$_{26.5}$ isophotal magnitudes for 6756 
galaxies in a region covering 60~$\times$~25~arcmin$^2$ in the center 
of the Coma cluster, plus 267 galaxies in a region of
9.7~$\times$~9.4~arcmin$^2$ around NGC~4839,
we derive the luminosity function in the magnitude range 
13.5$\leq V_{26.5} <$ 21.0 (corresponding to the absolute magnitude range
$-22.24 < M_{V26.5} \leq -14.74$). The luminosity function for this 
region is well fitted by the combination of a gaussian in its bright 
part and of a steep Schechter function (of index 
$\alpha =-1.8$) in its faint part. Luminosity functions derived for
individual regions surrounding the brightest galaxies show less steep
slopes, strongly suggesting the existence of environmental effects.
The implications of such effects and galaxy formation scenarios are
discussed.

\end{abstract}

\section{Introduction}\label{intro}

The shape of the luminosity function (LF) of galaxies gives strong constraints 
on cosmological parameters and formation scenarios since it is closely related 
to the galaxy mass function (MF) and therefore to the spectrum of initial 
perturbations (see Binggeli et al. 1988 for a review). For instance,
 the hierarchical model predicts a MF characterized by
an exponential cut-off above a given mass ($M^*$) and a power law 
(with index $\alpha$) at low masses, where the index $\alpha $
is related to the slope n of the power spectrum as $\alpha=(9-n)/6~$ 
with $-3<n<0$ (Schechter 1976). Such a typical behaviour has been 
found in the studies of the LF both for field or group and cluster 
galaxies, and has been popularized as the ``Universal Schechter function''.

However this natural link between LF and MF is probably somewhat simplistic
since environmental effects expected to be present at least in high density 
regions like clusters of galaxies have been neglected and could probably play
 an important role in modifying morphologies, luminosities, etc ...

In the field, recent spectroscopic surveys (Zucca et al. 1995, 
Ellis et al. 1996) lead to a rather shallow
 slope ($\alpha  \simeq -1 $) for the faint luminosity part. For 
clusters of galaxies, the situation is more intricate. 
There seems to be nevertheless a consensus that in such systems the slope of 
the LF is steeper than in the field. Recent works by Bernstein et al. 
(1995, hereafter BNTUW) and De Propris et al. (1995) give  values 
ranging from $\alpha = -1.4$ to the extreme value of $-2.2$. 

Such large values seem to indicate that a numerous  population of faint 
galaxies is present in clusters and not in the field. Such a segregation 
could be related either to the formation conditions
(``nature'' effects, where denser initial density peaks leading to clusters 
fragment into smaller subunits than in the field) or to 
environmental effects (``nurture'' effects, where e.g. the intra-cluster gas 
would be able to confine small cluster galaxies, which would not be the case
in the field).

Comprehensive comparative determinations of LFs in the 
field and in clusters as well as in various environments inside 
clusters will clearly give valuable clues on the above topics. 

Such studies are generally restricted by insufficient observational 
material. Either the field areas studied are small in order to sample enough 
objects below  $ M^*$ (the faint end of the LF), or, on the contrary, the 
magnitude range obtained is limited when observations are made on large 
fields in order to increase the number statistics.

Last but not least, background subtraction plays an important role and
can in general be treated only statistically since redshifts are 
not available for the faintest objects.

In this paper, we present a survey of the Coma cluster performed at CFHT 
covering a region of 60 x 25 arcmin${^2}$, 
plus a second region of 9.7~$\times$~9.4~arcmin$^2$ around NGC~4839,
in which  V$_{26.5}$ 
magnitudes have been obtained for 6756 and 267 galaxies respectively, down to 
V$_{26.5}$=22.5, and address these questions.

In section~\ref{data} we briefly describe the data thus obtained and discuss 
the question 
of background subtraction. In section~\ref{fdl} we present the results for 
the LF 
in various regions of the cluster. Finally, in section~\ref{discussion} we 
discuss these 
results, in particular the steep slope at the faint end and the consequences 
of the dynamical history of the cluster on the LF.

\section{The data} \label{data}

\subsection{Galaxy magnitudes}\label{mag}

The results presented below are based on a CCD photometric catalogue 
of 6756 galaxies, obtained with a mosaic of images covering a region 
of 60~$\times$~25~arcmin$^2$, equivalent to 
2.4~$\times$~1~h$_{50}^{-2}$~Mpc$^2$,
 centered on the two brightest central 
galaxies of Coma (NGC~4874 and NGC~4889), plus one field of 
9.7~$\times$~9.4~arcmin$^2$ around NGC~4839 (the group south-west of 
Coma) with 267 galaxies. The latter region was treated separately and was not included 
in the overall luminosity function derived in section~\ref{allfdl}. All the 
details concerning the observations and data reduction are given in a 
paper by Lobo et al. (1996).  In the 
present study, the quantitites of interest taken from this catalogue 
are the coordinates and V$_{26.5}$ isophotal magnitudes for each 
galaxy.

We have checked that magnitudes computed inside the isophote 26.5 are 
optimal for our observational conditions and closely approximate total 
magnitudes. 
V$_{26.5}$ magnitudes will simply be noted as V hereafter.

Absolute magnitudes were 
computed with a distance modulus of 35.74 adopting a flat cosmology 
(q$_0$~=~1/2, $\Lambda$~=~0, and Hubble's constant 
H$_0~=~50$~km~ s$^{-1}$Mpc$^{-1}$) and taking into account a 
K-correction of 
0.03 magnitudes to all galaxies. 
The turnover of the counts is observed at V $\sim$ 22.5, but there is most 
probably also a
loss of completeness for galaxies brighter than 22.5, as discussed e.g. by 
BNTUW. 
Note that our central surface brightness limiting detection value of 
$\mu _0~=~24$ mag/arcsec$^2$ might have missed some very faint surface 
brightness objects (this value was simply determined {\it a posteriori} 
by the standard analysis of the V vs $\mu_0$ plot and a closer look
at the objects with fainter $\mu_0$.)

\begin{figure}
\centerline{\psfig{file=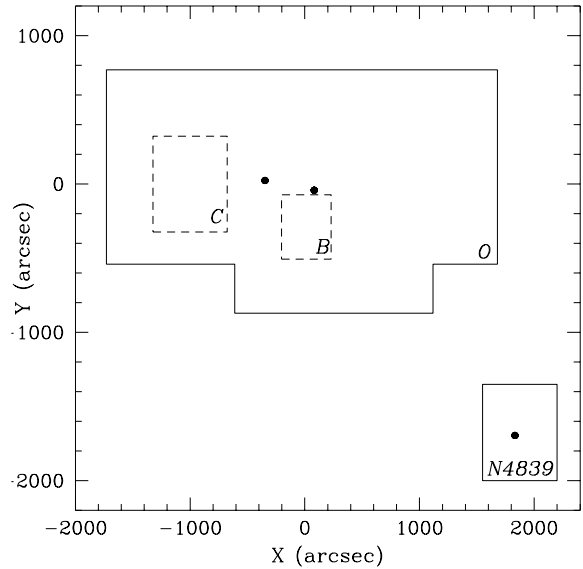,height=8.0cm}}
\caption{Map of some of the regions where we computed the luminosity function. 
Main regions are labeled, and the dots indicate the positions of the big 
galaxies NGC~4874 and NGC~4889 in the centre, and NGC~4839 in the south-west 
frame. Coordinates are given relatively to the GMP centre located at 
$\alpha=12^h\ 57.3^m$, $\delta =28^\circ\ 14.4$' (1950.0). North is up, east 
is to the left.}
\protect\label{carte}
\end{figure}

\subsection{Subsamples} \label{sub}

In our search for possible environmental effects, we also derived 
luminosity functions for different smaller spatial zones: one square 
region of 52.2~arcmin$^2$ coinciding exactly with that analyzed by 
BNTUW in their analysis of the deep luminosity function in the R-band, 
two square 
regions covering approximately the same area but this time centered on 
the bright 
galaxies NGC~4874 and NGC~4889, which have recently been shown to be 
the centers of groups (see Biviano et al.  1996), a third square 
region covering the same area, containing no bright galaxy and offset 
from the cluster center to the south-east by ($-1000,0$) arcsec, and the 
region of 
9.7~$\times$~9.4~arcmin$^2$ around NGC~4839 defined above.  These five 
regions will be labeled {\it B}, {\it N4874}, {\it N4889}, {\it C} 
and {\it N4839} respectively while the Overall region will be labeled {\it O} 
(see Fig.~\ref{carte}).

\subsection{Background subtraction}\label{back}

Field subtraction is never straightforward and the fact that we only 
possess observations in the cluster region (no additional frames of 
regions devoid of groups or clusters were obtained during the run) and 
one filter only, complicates even further this task.  We therefore 
decided to divide the sample into two subsamples according to 
brightness.

For bright galaxies (V$<18.5$), we first identified 
cluster members in our V catalogue by matching positions (to better than 
5 arcsec) with the Godwin et al. (1983, hereafter GMP) catalogue. Then we 
identified the galaxies as cluster members, either 
by redshift or by location in the colour-magnitude band defined 
by Mazure et al. (1988).  A more detailed description of this 
procedure can be found in Biviano et al. (1995).  Apparent 
magnitudes were converted to absolute ones by subtracting the distance 
modulus for Coma. Since the \CMB criterion is reliable at least up to 
b~=~19.0 (as tested by redshift information, see Biviano et al. 
1995), we decided to consider this subsample to be fairly correct up 
to this value, which typically corresponds to V $\sim$ 18.  At this 
stage we have 309 cluster members; field galaxies in this magnitude 
range constitute $\sim 25 \%$ of the V catalogue in the region of 
observations.
\medskip

For intermediate to faint galaxies (V $>$ 18), 
redshift determinations are rare and extrapolation of the \CMB is 
highly speculative.  We used the imaging catalogue of field galaxies 
from the ESO-Sculptor Faint Galaxy Redshift Survey (Bellanger {\it et 
al.} 1995, Arnouts et al. 1996, hereafter also noted as Sculptor 
data) to estimate the number of fore $+$ background objects per 
magnitude bin.  In this survey, observations were made for more than 40 
fields covering a total area of 1440~arcmin$^2$ with seeing conditions 
similar to those of our run.  Total V magnitudes were obtained for
 a complete sample of objects ranging from V~=~18.0 
to V~=~23.5.  We built a mean field with these counts and scaled it to 
the area covered by our observations, obtaining a sample of 3782 field 
galaxies in the range $18.0 < V \leq$ 22.5.  Symmetrical error bars 
for each magnitude bin take into account
fluctuations among the different frames.  We then subtracted these 
field counts to the corresponding magnitude range of our V catalogue.  
This approach is valid since the Coma cluster is essentially free of 
internal extinction (Ferguson 1993), and the galactic reddening is 
negligible in the direction of Coma (Burstein $\&$ Heiles, 1982).  The 
subtraction of the field counts was made directly bin per bin, after 
binning the Sculptor and the Coma V catalogues in exactly the same way 
with 0.5~mag/bin. We then converted apparent magnitudes to absolute 
ones. Cluster member galaxies add up to 884 for $18.0 < V \leq$ 21.0 
($-17.74 <$ M$_V \leq -14.74$), and field galaxies represent 47.3$\%$ 
of the observed initial sample.\\
\begin{figure}
\centerline{\psfig{file=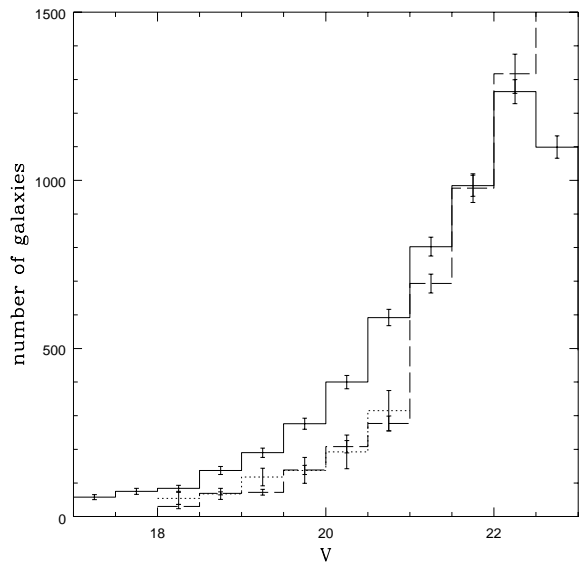,height=8cm}}
\caption{Raw counts derived from the observations (continuous line), 
background ESO-Sculptor counts (long-dashed line) and luminosity function 
for Coma after field subtraction (dotted line) in the Overall region 
{\it O} for V$\leq 21.0$ (see text).}
\protect\label{counts}
\end{figure}
For the faintest magnitudes (V$>21.0$), it can be seen from 
Fig.~\ref{counts} that problems do exist for field estimation, as the field 
counts exceed those in the region of observations. 
This causes a decrease towards negative values of the luminosity function 
(after field subtraction). This 
phenomenon is most probably due to the fact that the Coma sample is 
incomplete at these magnitudes, therefore making the background 
subtraction hazardous.  We will therefore limit our analyses to V$\leq 
21.0$. The resulting error bars for Coma data after field subtraction 
were computed in each magnitude bin using the classical formula: 
$\sigma_C^2=\sigma_{all}^2 + \sigma_f^2$, where subscript {\it C} 
stands for Coma data after field subtraction, {\it all} for the 
observed field plus Coma galaxy number distribution, and {\it f} for 
field galaxies.  $\sigma_{all}$ is assumed to be poissonian, so we 
take it to be equal to $\sqrt{N_{all}}$; $\sigma_f$ is the error bar 
for each magnitude bin as computed by Arnouts {\it et al} (1996).
\begin{figure}
\centerline{\psfig{file=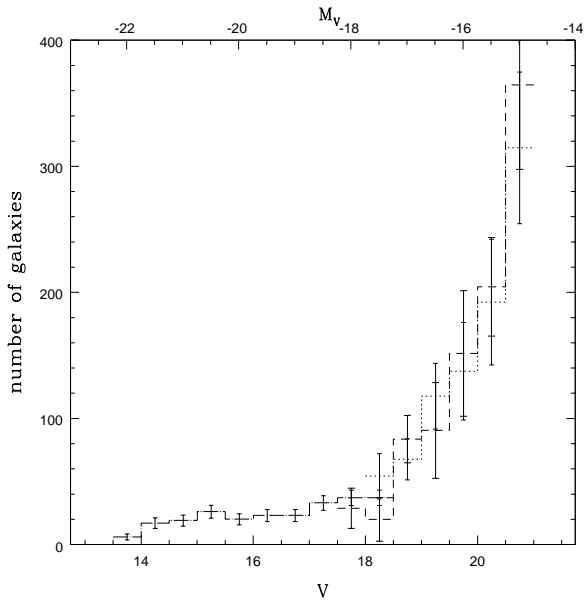,height=8cm}}
\caption{Luminosity functions derived for Coma in the Overall region: the 
bright part (dot-dashed line) is superposed to the fainter ones: 
respectively, that obtained after subtraction of the CFRS field counts 
(dashed line) and that derived with the background provided by the 
ESO-Sculptor survey (dotted line).}
\protect\label{superposed}
\end{figure}
The result obtained by direct bin subtraction is questionable due to binning effects. However, we have tested and confirmed its 
reliability by fitting the field counts with mathematical functions and then 
subtracting this fit to our observations in order to discard any possible 
irregularities particular to the background fields in question. The result 
was the same, within error bars, as that obtained after direct subtraction.
To strengthen our confidence in these results we also performed the 
same general process of field subtraction using the field counts 
derived from the Canada-France Redshift Survey (CFRS, Lilly {\it et 
al.} 1995a) in a field of 400~arcmin$^2$, which have the advantage of 
having been made with the same instrument as ours and under similar 
conditions.  We indirectly obtained isophotal V magnitudes, 
V$_{iso} \geq 17.5$, by means of their observed I$_{iso}$ ({\it iso} being a 
sufficiently low isophote to be sure that isophotal magnitudes thus 
derived are 
very close to total magnitudes and to our own 26.5 isophote), and of 
3'' aperture V and I magnitudes.  The luminosity function we obtained 
for Coma after scaled subtraction of the CFRS counts was consistent 
with that computed with Sculptor (see Fig.~\ref{superposed}).  However, 
as the error bars we 
computed - following the same recipe as above - for the V$_{iso}$ 
magnitude bins from the CFRS are larger than the Sculptor ones (because 
the CFRS surveyed area is smaller) we 
prefered to continue our analyses with the luminosity function 
obtained by subtracting the Sculptor counts.
\medskip
For the whole magnitude range (13.5 $\leq$ V $\leq$ 21.0), when we compare the 
number of galaxies in the brighter range to that in the faint range, we find 
compatible values in the overlap zone (i.e. the bins centered on 
V~=~17.75 and 18.25), 
within error bars. This ``junction condition'' will play an important 
role in the discussion of the quality or the robustness of the results 
presented in section~\ref{quality}.  We therefore chose to adopt for the luminosity function the 
values determined from the \CMB criterion up to V~=~18.5 (included), as 
this method is less sensitive to number count fluctuations in the 
extreme bins.

Note that statistical handling of contamination by foreground and 
background objects, as we have done it, is allowed due to the high density 
of the cluster core.  However, the procedures described for 
eliminating fainter ``intruders'' are tricky and errors are expected 
from background fluctuations, specially at the faint end (where we 
seem to have somewhat overestimated the field number counts), as discussed 
by Colless (1989).  Apart from several possible problems of background 
estimation, we nevertheless confide in the robustness of our result: 
the simple fact that we obtain the same luminosity distribution when 
using two different field galaxy surveys seems to indicate that our 
background subtraction is correct.

We therefore obtain a continuous magnitude distribution of 1074 objects 
in the range V$\leq$ 21.0, corresponding to the absolute magnitude range
M$_V \leq -14.74$, which we shall now use to draw and
characterize the luminosity function.

\section{The luminosity functions}\label{fdl}

\subsection{The overall luminosity function}\label{allfdl}

The luminosity function in the b-band was recently studied in detail 
by Biviano et al. (1995), who fit their data with the following 
functions and/or their combinations: 1)~a Schechter function (S), 
2)~a Gauss function (G), and 3)~a Gamma distribution (also called Erlang 
function). In this paper we will fit our V data by a 
Maximum-Likelihood technique (following the method described in 
Bevington \& Robinson 1992) with the following functions:
 \begin{equation} 
S(M_V)=K_{S} \, 10^{0.4(\alpha+1)(M_V^{\ast}-M_V)} 
\exp[-10^{0.4(M_V^{\ast}-M_V)}] 
\end{equation} 

\begin{equation} 
G(M_V)=K_{G} \, \exp[-(M_V-\mu)^{2}/(2\sigma^{2})] 
\end{equation}

\noindent and by their combination.

\begin{table*}
\caption[ ]{Different fits to the whole luminosity function.}
\begin{tabular}{cccccccc}
\hline\hline
Function & Mag range & $M_V^{\ast}$ & $\alpha$ & $\mu$ & $\sigma$ & 
$\chi^2_R$\\
& for the fit & & & & & & \\
\hline
S & $-22.24<M_V\leq~-14.74$ & $-29.0\pm$2.0 & $-1.59\pm$0.02 & & & 
3.1 \\
S + G & $-22.24<M_V\leq~-14.74$ & $-22.7\pm$0.4 & $-1.80\pm$0.05 & 
$-20.4\pm$0.2 & 1.1$\pm$0.3 & 0.6\\
\hline
\end{tabular}
\label{table1}
\\ \\ \noindent
G=Gauss, S=Schechter
\end{table*}
\begin{figure}
\centerline{\psfig{file=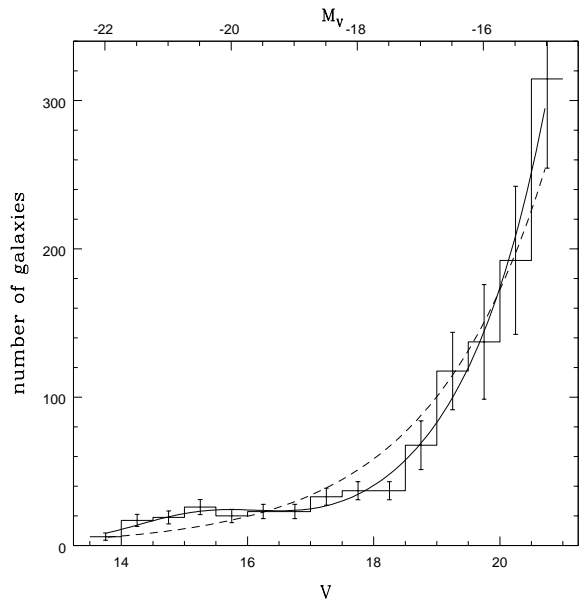,height=8cm}}
\caption{Schechter (dashed-line) and Schechter + Gauss (continuous line) fits 
to all the Coma data in the Overall region. 
}
\protect\label{fitSG}
\end{figure}
We give in Table~\ref{table1} the best fit results obtained after 
excluding the three brightest cluster members. The reason for this 
has been discussed by Biviano et al. (1995), and is essentially due to the
influence of these three galaxies
on the shape of the gaussian. This condition limits the data to the 
range $-22.24 <$ M$_V \leq ~-14.74$.  The columns of 
Table~\ref{table1} are the following: (1)~function used to fit the 
luminosity function, (2)~magnitude range for the fit, (3) to (6)~values of 
the best fit parameters, (7)~reduced chi-square.

It can be seen from this Table and Fig.~\ref{fitSG} that a single Schechter 
function does not fit the data.  On the other hand, the combination of a 
Schechter and a Gaussian fits the data quite well.

We have also tried to fit the 
luminosity function with the combination of a Schechter and an Erlang 
function (see Biviano et al. 1995).  The 
Maximum-Likelihood-Ratio test (see e.g.  Meyer 1975) shows that both G 
and E functions fit the data equivalently, so we will therefore limit 
our discussion to the G function.
We can note that the value of $M^*_V$ is well constrained and the 
slope of the Schechter function is quite steep: $\alpha~=~-1.8$.  The 
characteristics of the Gaussian are quite similar to those found in the 
b-band by Biviano et al. (1995), assuming a color index $(b-V)$=0.9.  This 
will be discussed in section~\ref{fdlbv}.

\subsection{Luminosity function in several subsamples}\label{subfdl}

BNTUW recently addressed the same question in a much smaller region of Coma, 
but using deeper photometry taken with a different filter (R). In particular, 
they fitted the luminosity function over the range 15.5$<$R$<$23.5 
(approximately equivalent to 16.5$<$V$<$24.5 if we take a typical value
$(V-R)$=1.0 for cluster 
galaxies).  Parametrizing it as a power law dN/dL~$\propto$ L$^\alpha$, they 
derive $\alpha~=~-1.42 \pm 0.05$ ($\chi^2_R$~=~0.8), and their 95$\%$ 
confidence interval is $-1.57 <$ $\alpha<$ $-1.25$. The fact that we found 
a spectral index for the Coma luminosity function notably steeper than 
that estimated by BNTUW worried us at first, because we expected to observe
the same stellar populations in V and R.

This led us to estimate the luminosity function for region {\it B}, and
also for a few sub-regions (defined in section 
\ref{sub}) characterized by obvious peculiar features, in exactly the
same manner as BNTUW. All these 
luminosity functions were fitted in the interval 16.5$<$V$\leq$21.0 by 
the same power law as used by BNTUW.  Results are given in 
Table~\ref{table2}. 
\begin{table}
\caption[ ]{Power law fits to the data in the magnitude interval
$16.5<V\leq~21.0$ ($-19.24<M_V\leq~-14.74$) for different regions.}
\protect\label{table2}
\begin{tabular}{lcrr}
\hline\hline
Region & Power law slope $\alpha$ & $N_{gal}$ & surface area \\
       & $\pm1\sigma$ error          &           &$arcmin^{2}$~~  \\ 
\hline
{\it O}     & $-1.81\pm$0.03 & 1025& 1422~~~~\\
{\it B}     & $-1.71\pm$0.11 &   60& 52~~~~\\
{\it N4874} & $-1.58\pm$0.10 &   57& 51~~~~\\
{\it N4889} & $-1.51\pm$0.13 &   32& 51~~~~\\
{\it N4839} & $-1.74\pm$0.11 &   54& 117~~~~\\
{\it N4839} & $-1.79\pm$0.13 &   38& 51~~~~\\
{\it C}     & $-1.88\pm$0.10 &   77& 116~~~~\\
\hline
\end{tabular}
\end{table}

If we take the magnitude distribution of the {\it O} sample as the parent 
distribution for all subsamples, we can test the null hypothesis that the 
magnitude distributions of these subsamples are drawn from this parent 
distribution. We parametrize the magnitude distributions by their 
$\alpha$ indices; assuming that the errors on these indices are normally 
distributed around the observed values, we give in Table~\ref{table3} the 
probabilities that the $\alpha$'s are the same as that derived for the 
reference sample {\it O}.
 
\begin{table}[tbp]
	\caption{Probabilities for the $\alpha$ indices obtained in various
        subsamples to be the same as the $\alpha$ index of the parent
        sample~{\it O}.	}
	\protect\label{table3}
	\begin{tabular}{ccccc}
		\hline\hline
	{\it C} & {\it N4874} & {\it N4889} & {\it B} & {\it N4839}  \\
		\hline
0.38 &0.04 & 0.04 & 0.38 & 0.95\\
		\hline
	\end{tabular}
\end{table}
\medskip
 The results in Tables~1-3 lead to the following comments:
\begin{itemize}
	\item  The $\alpha$ index of the power law in the overall cluster 
($-1.81$) is consistent with the index of the Schechter function derived 
previously (see Table~\ref{table1}). 

	\item  The index found in region {\it B} is much steeper than that of 
BNTUW (see discussion below). 

	\item   The power law index found for regions {\it N4874}
and {\it N4889} is significantly flatter, while for region {\it N4839} it is
identical to that computed for {\it O}.

\end{itemize}

These results could be due to the 
environmental properties of regions {\it N4874}
and {\it N4889}, and could also be linked to other properties 
described in a previous paper (Biviano et al. 1996). However, the flatness
of $\alpha$ is probably not simply
 due to the existence of groups around these two galaxies, since such a flat
index is not observed in the {\it N4839} region.
As a comparison, we have also derived $\alpha$ in a region 
without peculiar features (region {\it C}~) but with a similar number of 
galaxies, and we find a value that does not differ very much from the overall 
index. We will discuss the hypothesis of environmental effects and its 
implications in section~4.3.

\subsection{Quality of our results}\label{quality}

First of all, we emphasize the fact that we have done our analyses and 
fits using a large number of galaxies both in the central part of Coma 
and in the two background samples. 
We give below several arguments justifying the robustness of our results.

The fact that subtracting the background contribution 
estimated from two different samples (Sculptor and CFRS) leads 
to similar results strongly suggests that our background 
subtraction is correct. However, various kinds of effects could affect
this background subtraction. 
First, we could have made some mistakes when comparing our V 
magnitudes to those derived by the Sculptor survey team. We can test the 
effect of such errors by subtracting the background after shifting it by 
V~=~$\pm 0.5$, and performing again our Schechter plus Gauss fit. 
The $\alpha$ indices resulting from such fits on the {\it O} sample
turn out to be $-1.79\pm$0.03 and $-1.48\pm$0.05, as compared to
$-1.80\pm$0.05 (Table~1). However, in both cases, the ``junction 
condition'' (defined in section~\ref{back}) is badly fulfilled; moreover, 
$M^{*}$ is out of the magnitude range used in the fitting procedure. 
$\alpha= -1.48$ is a limiting value allowed by our data; however, it still 
represents a steeper slope than that observed for field galaxies. In addition, 
we checked that our magnitudes do not differ significantly 
from V$_{Kron}$ (the Sculptor data actually consists of Kron magnitudes).

Second, it could also be possible that the actual background of Coma 
differs from what we consider as a universal one, but unless we obtain 
redshifts for all the galaxies in the Coma area, this cannot be checked.  
We will try to approximate such a local variation in two ways:
\begin{enumerate}
	\item  There can be effects due to the total number of galaxies 
subtracted, the slope of the background remaining constant. 
Again, the ``junction condition'' suggests that this 
first effect is certainly weak.
	\item There could be a variation of the background slope, with the 
``junction condition'' still being fullfilled.  Notice that BNTUW have 
discussed and simulated the variation of data completeness from bin 
to bin.  If the effective background slope derived from the Sculptor data 
has been overestimated, the final $\alpha$ index will be flatter 
than the real one, and this effect will be superimposed to that due to 
the incompleteness of the sample. In this case, the value of $\alpha 
=-1.8$ is an upper limit (algebraically). On the other hand, if the 
background slope has been
underestimated, both effects will be antagonistic, and no definite conclusion 
can be drawn.
\end{enumerate}

With the likely hypothesis that our background subtraction is correct, we now 
turn to the comparison between our results for region {\it B} and 
those of BNTUW. They have estimated $\alpha= -1.42 \pm 0.05$ while we 
have found: $\alpha= -1.71 \pm 0.11$ (see Table~\ref{table2}).
However, the range of magnitudes of their analysis ($15.5<R<23.5$) is much 
larger than ours ($16.5<V\leq21$, assuming $V-R \sim 1$).

\begin{figure}
\centerline{\psfig{file=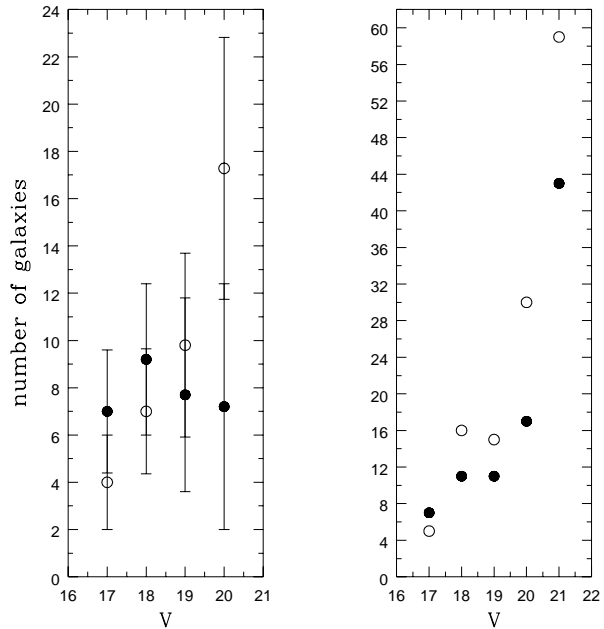,height=9cm}}
\caption{Comparison between Bernstein et al. (1995) data (filled 
circles) and our own (open circles) in the {\it B} region. Left panel: 
luminosity functions. Right panel: raw counts}
\protect\label{bernstein}
\end{figure}

We display BNTUW's data (their Table~2) and ours on the left panel of 
Fig.~\ref{bernstein}, as well as the raw counts on the right panel of the same 
figure.  
There is clearly a large discrepancy in the two samples, even in the 
bright bins where the same galaxies should be observed in V and R. 
Notice that for the range $18<V<20$, the data indicated by BNTUW show a very 
irregular behaviour and would lead to a flat slope; it is only the data for 
larger values of V which allow them to find $\alpha= -1.42 $. 

One cannot tell if the discrepancy between our slope in the {\it B} region and
that found by BNTUW is due to the $(V-R)$ transformation. However, we have 
confidence in our result, since the $\alpha$ index 
that we obtain differs by less than one $\sigma$ from 
that derived for the {\it O} region. {\it B} is a fair 
sub-sample, representative of the parent sample, even if the 
number of galaxies belonging to it is small. 

All the reasons presented in this section show the reliability of the 
{\it Overall} result, thus reinforcing our confidence in the result 
obtained for the {\it B} sub-sample.

\section{Discussion and Conclusion}\label{discussion}

\subsection{Steep spectral index}\label{index}

The steep Schechter index derived above for the 
intermediate to faint galaxies ($\alpha =-1.8$)
 suggests the existence of an excess dwarf 
population compared to the field, where the slope is rather $-1.3 < 
\alpha< -1.1$ (see e.g. Zucca et al. 1995, Ellis et al. 1996, and 
references therein).  Recently, deeper counts from HST data have been 
performed by Driver et al. (1995); but these authors have shown 
that models using a Schechter function with $\alpha~=~-1.8$ cannot 
really account for the observations.

This steep index is in clear contradiction with previously derived values
for both groups and clusters (e.g.  Ferguson \& Sandage 1991), and for Coma
in particular (Thompson $\&$ Gregory 1993), which 
were all much flatter. However, several other observations of galaxy clusters 
show an abrupt rise in the faint end of the luminosity function, thus
revealing the presence of a large number of faint galaxies.
A value of $\alpha~=~-1.8$ was 
determined by Driver et al. (1994) for a z=0.2 cluster 
($-24 < M_R < -16.5$). De Propris et al. (1995) derive $\alpha~=~-2.2$ 
for the cores of four rich 
nearby clusters (A2052, A2107, A2199 and A2666) in the approximate magnitude 
ranges  $-15~<~M_B~<~-11$ and $-16~<~M_I~<~-13$.  
These authors also present a complete discussion and revision of the 
values of the Schechter index obtained for other clusters and for the 
field. 

 In fact, the value of the Schechter index clearly depends on 
the part of the luminosity function which is being studied: the faint part 
of a cluster luminosity function is steeper than the bright part.  BNTUW 
also found a dramatic upturn at M$_R=-11.9$ (R=23), 
where the steep (dwarf) component takes over.  This change of slope is in 
contradiction with what has been found up to now for field photometric 
surveys (Lilly et al. 1995b, Ellis et al. 1996).

The over-abundance of dwarf galaxies in clusters is an observational result 
that one has to take into account in the 
general theory of segregation according to galaxy density, such as the 
well known segregation by type, etc... Babul $\&$ Rees (1992) suggest 
that dwarf galaxies ``will be strongly influenced by the local intergalactic 
medium; those 
	in high-pressure environments [the intra-cluster medium] will be 
	preserved while those in low-pressure ones [the field] will fade 
away'', their gas being easily ejected by supernova driven winds. 
This is an idea very close to the concept that the bias parameter depends 
on the local density (Gnedin 1996). Explaining why there is an over-abundance 
of dwarf galaxies is however beyond the scope of this paper.

\subsection{Double distribution of luminous galaxies}\label{fdlbv}

If we consider the luminous galaxies (up to V$\sim$17), part of them follow a 
Gauss distribution while 
the others are distributed according to the Schechter function
previously determined. One can therefore say that, at least as far as
their luminosity properties are concerned, there are two different 
populations of luminous galaxies.  

We have compared the results described above to those obtained for the b 
magnitudes of the same objects by Biviano et al. (1995), using a 
typical colour index ($b~-~V \sim 0.9$).

Before making any other comment, one must first notice that the depth of 
these two analyses is different: b$<20.0$ versus V$\leq 21.0$, which roughly 
corresponds to b$\leq 22$. However, we also stress that for the two analyses 
we have found that galaxies are distributed in a composite manner, 
following a gaussian plus a Schechter distribution.  When comparing 
the V and b luminosity functions, one should expect the fainter part to be 
better constrained in the V-band, where photometry goes deeper, and this is 
the case. 
Furthermore, we confirm the excess of bright galaxies (well 
fit by a gaussian in this work) followed by a flattening 
already apparent in the works of Rood (1969), Godwin $\&$ Peach (1977), Lugger 
(1986), Thompson $\&$ Gregory (1993), Kashikawa et al. (1995), and 
well defined in Biviano et al. (1995).

However, the characteristics of the two fits are not identical: 
it has not been possible to find the same gaussian fit 
in V and in b (that is, same central magnitude and dispersion) 
simultaneously with identical values for the 
Schechter distribution parameters, even in equivalent magnitude ranges 
(b$<20.0$ and V$<19.0$). 
However, the mean value $\mu$ of the Gauss distribution is 
(within error bars) generally the same. 
These results are in agreement with the eye impression: the dip in b 
appears far more obvious than in V, where the bump is apparently more 
``diluted''; notice that there is a slight indication that the $\sigma$ 
value given in Table~\ref{table1} for the S + G fit is larger than the 
$\sigma$ given in Table~2 of Biviano \textit{et al} (1995). 
However, correlation tests clearly show that the b 
and V distributions are highly correlated in the bright range 
(M$_V \leq -17.24$): both parameters $\tau$ of Kendall and $\rho$ of 
Spearman (see e.g. Kendall \& Stuart 1977) are close to 1.0 (0.866 and 0.971, 
respectively).\\

This difference between the b and V data would suggest the existence of 
blue galaxies (highly populated with young stars) outstanding in the b 
filter observations but being missed in V.  To test this hypothesis it 
would be useful to have infra-red photometry or, alternatively, 
ultra-violet data.  Quite recently, Donas et al. (1995) 
presented UV photometry for the Coma cluster and proposed the idea 
that a starburst event has taken place less than 1~Gyr ago.  By 
the galaxy colour indices (m$_{UV}-b$) they infer that residual star 
formation is still present in some of them; these galaxies also show 
peculiar spectra with emission lines indicative of enhanced star 
formation in the past or still in course (Caldwell et al. 1993); 
these lines are surprisingly strong for their morphological 
classification (early-type E-S0).  Bluer passbands are the most 
sensitive to star formation signatures, so this kind of burst would be 
sufficient to shift some of the galaxies in Coma towards brighter 
magnitudes in b, affecting V magnitudes only slightly and thus causing 
the different behaviours of the luminosity functions in b and V.  For 
the moment, this scenario is only speculative and we intend to 
investigate further this subject in future works.

\subsection{Environmental effects}\label{environ}

Table~\ref{table2} shows that the $\alpha$ indices are steep in the overall 
cluster and region {\it C}, while flatter for the two groups 
{\it N4874} and {\it N4889}, contrary to {\it N4839}. 
We have already shown that the regions surrounding NGC~4874 and NGC~4889 
do have particular properties, in 
particular a kernel map has revealed that they were both surrounded by 
an excess of bright galaxies (Biviano et al. 1996); since the $\alpha$ 
index we compute is that characterizing the intermediate and faint 
galaxy populations, this implies a lack of faint dwarf galaxies in these two 
regions. This result suggests an environmental effect in the cluster 
core correlated to the presence of these two substructures.

The mere presence of these two giant galaxies obviously implies that 
the luminosity functions in the regions surrounding these two galaxies
differ from that in any other zone of the cluster, at least in the bright
part. Since the power-law fit is made in a magnitude interval located
far from the magnitudes of these two galaxies, the slopes we have found in
regions {\it N4874} and {\it N4889} do indicate that the luminosity functions
in these regions really show a different behaviour.

Although the number of faint galaxies in these two regions is very high, 
the corresponding luminosity is very small: 76$\%$ of the number of 
galaxies in the {\it N4874} area represent only 5.5 $\%$ 
of the total luminosity, while the four most luminous ones account for 
more than 50$\%$. It is also interesting to notice that if we 
compare the light distribution in regions {\it N4874} and {\it N4889} to that 
in the {\it O} region, we find that there is more light (i.e. larger galaxies) 
in the luminous part, followed by less light (i.e. small galaxies) in the 
faint range, just as if a balancing process was in progress.

The relative lack of 
faint galaxies in regions {\it N4874} and {\it N4889} might be 
linked to the recent history of these two subclusters. It is generally 
believed (see e.g. Merritt 1985) that the cD (or D) galaxies result from a 
cannibalism process occuring in a group; the subsequent infall of the group
 into a cluster gives it the appearance of a sub-cluster.  The gravitational 
field of these bright galaxies in the group is locally so important 
that a large number of very small or dwarf galaxies 
must be satellites of the brightest ones, i.e. must be more strongly linked 
to the bright galaxies than to the group as a whole.  
We can speculate that when the group 
falls into the central region of a large cluster, the clouds of 
satellites are strongly affected in their trajectories by the tidal field 
and this could cause a rapid shower of dwarf galaxies towards the largest 
galaxies.

>From this point of view, we can understand the different index values of the 
{\it N4874} and {\it N4889} groups compared to that infered for the 
{\it N4839} group as due to a difference in their history.  
If the {\it N4839} group has not yet crossed the cluster, then the satellite 
shower has not yet happened. 
The history of this sub-cluster has been the target of controversial 
discussions between authors defending different scenarios (cf. 
Burns et al. 1994, Colless \& Dunn 1996). However, observational 
evidence seems to 
indicate that this group has actually crossed the main cluster (e.g. Caldwell 
et al. 1993) though not through the very central regions, as 
simulations show that tidal disruption would then have occured 
(Gonz\' alez-Casado et al. 1994; see also Biviano et al. 1996 for
 further details).
This scenario supports our interpretation of the cause for the different 
luminosity function slopes in the various groups.

\subsection{Summary of our results}\label{conclu}

We have firmly established (with good statistics) that the luminosity 
function in the central part of the Coma cluster shows the following 
characteristics:
\begin{itemize}
	\item[$-$]  The slope of the luminosity function is very steep, 
$\alpha~=~-1.8$, suggesting the presence of a huge number of faint and dwarf 
galaxies in the Coma cluster compared to the field. 
	\item[$-$] Although both b and V luminous galaxies are well 
	described by a composite distribution (a gaussian plus a Schechter
function), 
	their distributions are not identical. We therefore conclude that
from the luminosity point of view, there are two kinds of galaxies.
	\item[$-$] Luminosity functions derived for the two central 
sub-clusters present a relatively faint slope of $\alpha~=~-1.5$, suggesting
that there are fewer faint and dwarf galaxies in these regions than in the 
cluster as a whole. We have suggested a dynamical explanation for this 
observation, and because the luminosity function of the group around NGC~4839 
is the same as for Coma, it is also suggested that the history of this group 
is different from that of the two other sub-clusters. 
\end{itemize}

\begin{acknowledgements}
We are very grateful to St\'ephane Arnouts and Val\'erie de Lapparent
for providing us with the ESO-Sculptor V-counts and to the CFRS team for 
giving us access to their data. We thank Didier Pelat, Bob Nichol and Gary 
Bernstein for discussions, and Fred James for his help in the use of the
MINUIT software. 
We acknowledge financial support from GDR-Cosmologie, CNRS. CL is fully
 supported by the BD/2772/93RM grant attributed by JNICT, Portugal.
\end{acknowledgements}


\begin{thebibliography}{}
\bibitem {} Arnouts S., de Lapparent V., Mathez G., et al. 1996, A\&AS, 
in press
\bibitem {} Babul A., Rees M. 1992, MNRAS 255, 346
\bibitem {} Bellanger C., de Lapparent V., Arnouts S., et al. 
1995, A$\&$AS 110, 159
\bibitem {} Bernstein G.M., Nichol R., Tyson J.A., Ulmer M.P., 
Wittman D. 1995, AJ 110, 1507 (BNTUW)
\bibitem {} Bevington P.R., Robinson D.K. 1992, in {\sl Data 
Reduction and Error Analysis for the Physical Sciences}, 2nd ed. 
McGraw-Hill, p.~41
\bibitem {} Binggeli B., Sandage A., Tamman G.A. 1988, ARA$\&$A 26, 509
\bibitem {} Biviano A., Durret F., Gerbal D., et al.
1995, A$\&$A 297, 610
\bibitem {} Biviano A., Durret F., Gerbal D., et al.
1996, A$\&$A in press
\bibitem {} Burns J., Roettiger K., Ledlow M., Klypin A. 1994, ApJ 427, L87 
\bibitem {} Burstein D., Heiles C. 1982, AJ 87, 1165
\bibitem{} Caldwell N., Rose J.A., Sharples R.M., Ellis R.S., Bower
R.G. 1993, AJ 106, 473
\bibitem {} Colless M. 1989, MNRAS 237, 799
\bibitem {} Colless M., Dunn A.M. 1996, ApJ 458, 435
\bibitem {} De Propris R., Pritchet C.J., Harris W.E., McClure R.D. 
1995, ApJ 450, 534
\bibitem {} Driver S.P., Phillipps S., Davies J.I., Morgan I., Disney 
M.J. 1994, MNRAS 268, 393
\bibitem {} Driver S.P., Windhorst R.A., Ostrander E.J., et al. 
1995, ApJ 449, L23
\bibitem {} Donas J., Milliard B., Laget M. 1995, A$\&$A 303, 661
\bibitem {} Ellis R.S., Colless M., Broadhurst T., Heyl J., 
Glazebrook K. 1996, MNRAS 280, 235
\bibitem {} Ferguson H.C. 1993, MNRAS 263, 343
\bibitem {} Ferguson H.C., Sandage A. 1991, AJ 101, 765
\bibitem {} Gnedin N.Y. 1996, ApJ 456, 34
\bibitem {} Godwin J.G., Peach J.V. 1977, MNRAS 181, 323
\bibitem {} Godwin J.G., Metcalfe N., Peach J.V. 1983, MNRAS 202, 
113 (GMP)
\bibitem{} Gonz\'alez-Casado G., Mamon G.A., Salvador-Sol\'e E. 1994, ApJ 
433, L61
\bibitem {} Kashikawa N., Shimasaku K., Yagi M., et al. 
1995, ApJ 452, L99
\bibitem {} Kendall M., Stuart A. 1977, ``The Advanced Theory of Statistics'',
London, Griffin and Co. Ltd.
\bibitem {} Lilly S.J., Le F\`evre O., Crampton D., Hammer F., Tresse 
L. 1995a, ApJ 455, 50
\bibitem {} Lilly S.J., Tresse L., Hammer F., Crampton D., Le F\`evre O. 
1995b, ApJ 455, 108
\bibitem {} Lobo C., Biviano A., Durret F. et al. 1996, A\&AS submitted
\bibitem {} Lugger P.M. 1986, ApJ 303, 535
\bibitem {} Mazure A., Proust D., Mathez G., Mellier Y. 1988, 
A$\&$AS 76, 339
\bibitem{} Merritt D. 1985, ApJ 289, 18
\bibitem{} Meyer S.L. 1975, in {\sl Data Analysis for Scientists and 
Engineers}, ed. John Wiley \& Sons Inc., New York, p.~354
\bibitem {} Rood H.J. 1969, ApJ 158, 657
\bibitem {} Schechter P.G. 1976, ApJ 203, 297
\bibitem {} Thompson L.A., Gregory S.A. 1993, AJ 106, 2197
\bibitem {} Zucca E., Vettolani G., Cappi A. et al. 1995, Proc. 
International Workshop ``Observational cosmology: from galaxies to galaxy 
systems'', Sesto, Italy, July 4-7, 1995, G. Palumbo Editor
\end{thebibliography}
\end{document}